\documentclass[preprint,showpacs,preprintnumbers,amsmath,xamssymb,pra]{revtex4}

\usepackage{amssymb}
\usepackage{amsmath}
\input{epsf}

\begin{document}

\title{Slow electrostatic fluctuations generated by beam-plasma interaction}

\author{Karen Pommois, Francesco Valentini, Oreste Pezzi and Pierluigi Veltri}
\affiliation{Dipartimento di Fisica, Universit\`a della Calabria, 87036 Rende (CS), Italy}


\begin{abstract}
Eulerian simulations of the Vlasov-Poisson equations have been employed to analyze the excitation of slow electrostatic
fluctuations (with phase speed close to the electron thermal speed), due to a beam-plasma interaction, and their propagation in
linear and nonlinear regime. In 1968, O'Neil and Malmberg [Phys. Fluids {\bf 11}, 1754 (1968)] dubbed these waves ``beam modes''.
In the present paper, it is shown that, in the presence of a cold and low density electron beam, these beam modes can become
unstable and then survive Landau damping unlike the Langmuir waves. When an electron beam is launched in a plasma of Maxwellian
electrons and motionless protons and this initial equilibrium is perturbed by a monochromatic density disturbance, the electric
field amplitude grows exponentially in time and then undergoes nonlinear saturation, associated with the kinetic effects of
particle trapping and phase space vortex generation. Moreover, if the initial density perturbation is setup in the form of a low
amplitude random phase noise, once the most unstable mode has reached its nonlinear saturation amplitude after the linear growth,
the whole Fourier spectrum of wavenumbers is excited. As a result, the electric field profile appears as a train of isolated
pulses, each of them being associated with a phase space vortex in the electron distribution function. At later times, these
vortical structures tend to merge and, correspondingly, the electric pulses collapse, showing the tendency towards a time
asymptotic configuration with a single phase space structure associated to an electric soliton-like pulse. This dynamical
evolution is driven by purely kinetic processes, possibly at work in many space and laboratory plasma environments.
\end{abstract}

\date{\today}
\maketitle
 
\section{Introduction}
\label{sec:intro}
In 1946, Landau \cite{landau46} discussed the propagation of waves in unmagnetized plasmas and demonstrated that electrostatic
oscillations (Langmuir waves) of low amplitude can be damped in time, even in absence of collisional dissipation. The wave damping
process is triggered by the resonant interaction of the wave with particles moving with speed close to the wave phase velocity
$v_\phi$. Almost twenty years later, the Landau theory has been extended to the nonlinear regime by O'Neil \cite{oneil65}, who
showed that the nonlinear process of particle trapping in the wave potential can saturate Landau damping by flattening the
particle velocity distribution in the velocity range close to $v_\phi$: a Langmuir wave can survive Landau damping in the case its
amplitude is large enough that nonlinear effects come into play in a time much shorter than the Landau damping time. The
predictions by O'Neil have been confirmed in several numerical works
\cite{manfredi97,brunetti00,valentini05,galeotti05,galeotti06,califano06} and also in laboratory plasma experiments (see, for
example, Ref. \cite{danielson04}).

In 1991, Holloway and Dorning \cite{holloway91} noted that certain nonlinear electrostatic waves can propagate in a plasma at a
speed comparable to the electron thermal speed $v_{th,e}$, namely in a velocity range where they would be heavily Landau damped in
linear regime. They called these waves Electron Acoustic Waves (EAW), as their dispersion relation is of the acoustic form, and
showed that the presence of a population of electrons trapped in the wave trough turns off Landau damping. The excitation process
of the EAWs has been analyzed in numerical studies \cite{valentini06}, where it has been shown that an EAW is a
Bernstein-Green-Kruskal (BGK) \cite{BGK} nonlinear mode, and in laboratory experiments with nonneutral plasmas
\cite{kabantsev06,anderegg09a,anderegg09b}. In these works, the excitation of the EAW has been realized through an external driver
electric field of relatively low amplitude which, if applied for several nonlinear times, traps resonant electrons and flattens
the velocity distribution at $v_{th,e}$. Moreover, motivated by suggestions that EAWs might be launched in laser-plasma
interaction experiments \cite{afeyan04}, Jonhston et al. \cite{johnston09} focused on relatively large driver amplitudes. They
found novel BGK waves that they called “Kinetic Electrostatic Electron Nonlinear (KEEN) Waves.” These waves are comprised of four
or five significant phase-locked harmonics, persist only when driven hard enough, and are driven by a wide range of frequencies.

Recently, it has been shown \cite{valentini12} by means of kinetic numerical simulations that a slight deviation of the particle
velocity distribution, consisting of a small plateau of the width of a very small fraction of $v_{th,e}$, can give rise to new
undamped modes, named corner modes, that are the small amplitude limit of the nonlinear EAWs and can exist in a wide region of the
$k-\omega_R$ plane ($\omega_R$ being the real part of the wave frequency and $k$ the wavenumber). The corner modes are
intrinsically similar to the so-called beam modes, i. e. the electrostatic acoustic branch of waves arising from the interaction
of a background Maxwellian plasma and a relatively low density electron beam. The beam modes have been introduced in 1968 by
O'Neil and Malmberg, who analyzed the nature of the roots of the plasma dielectric function in presence of a beam \cite{oneil68}. 
The electron beam population, in fact, produces a slight deviation of the velocity distribution from Maxwellian, as for the 
corner modes, this allowing for the existence of an acoustic-like branch of electrostatic waves.

The interaction of a plasma with a particle beam is a physical phenomenon frequently observed in both space and laboratory 
plasmas and it is responsible for several dynamical processes, like, for example, the excitation of waves through instability 
mechanisms. In particular, in the solar wind, where the velocity distributions are routinely observed to deviate from the 
configuration of thermodynamic equilibrium \cite{marsch06} (with the appearance, for instance, of beams of accelerated 
particles), the presence of electrostatic slow waves with acoustic-like dispersion relation has been recently detected
\cite{valentini11a,valentini11b,vecchio14,valentini14b}. 

In this paper, we employ Eulerian simulations of the Vlasov-Poisson equations to investigate the linear and nonlinear regime of
propagation of the beam modes, in an unmagnetized plasma, composed of electrons and motionless protons. Firstly, we revisit the
linear results discussed in Ref. \cite{oneil68}, by using a linear kinetic solver which computes numerically the electrostatic
roots of the plasma dielectric function. Afterwards, we employ the fully nonlinear Vlasov-Poisson code to study the nonlinear
regime of the interaction of an electron beam, streaming at a speed close to $v_{th,e}$, with a background Maxwellian plasma.

This paper is organized as follows. In Section \ref{sec:lindisp}, we revisit the analytical, linear results described in Ref.
\cite{oneil68}. Therefore, in Section \ref{sec:numres} we present the numerical Vlasov-Poisson code used to investigate the
nonlinear system dynamics and discuss the numerical results. We finally summarize and conclude in Section \ref{sec:concl}.

\section{Linear dispersion relation of the beam modes}
\label{sec:lindisp}
As discussed above, the presence of a beam in the equilibrium distribution function influences the linear dispersion relation, 
as a new branch of acoustic modes, called beam modes, is generated. In this Section, by following Ref. \cite{oneil68}, we revisit
the linear dispersion theory of the beam modes. We focus on a collisionless plasma composed of kinetic electrons and a background
of neutralizing protons, in electrostatic approximation. Under these assumptions, the Vlasov-Maxwell set of equations can be
reduced to the 1D--1V (one dimension in physical space, one dimension in velocity space) Vlasov-Poisson (VP) system. In
dimensionless units, the VP equations read:
\begin{eqnarray}
\frac{\partial f}{\partial t}+v\frac{\partial f}{\partial x}-E\frac{\partial f}{\partial v}&=0 
\label{eq:vlas} \\  
 \frac{\partial E}{\partial x}=1-\int f\, dv \label{eq:pois}
\end{eqnarray}
where $f=f(x,v,t)$ is the electron distribution function and $E=E(x,t)$ is the electric field. In Eqs. 
(\ref{eq:vlas}--\ref{eq:pois}), time, lengths and velocities are scaled to the inverse electron plasma frequency
$\omega_{pe}^{-1}$, to the electron Debye length $\lambda_{De}$ and to the electron thermal velocity
$v_{th,e}=\lambda_{De}\,\omega_{pe}$, respectively. Moreover the electric field and the electron distribution function are
respectively normalized by $m_e v_{th,e} \omega_{pe}/e $ and $n_0/v_{th,e}$, $m_e$, $n_0$ and $e$ being the electron mass,
equilibrium density and electric charge, respectively. For the sake of simplicity, from now on, all quantities are scaled using
the characteristic parameters listed above.

The equilibrium electron distribution function, which is spatially homogeneous, has the following form:
\begin{equation}
 \label{eq:f0}
 f(x,v,t=0) = f_0 (v) = f_M(v) + f_b(v)
\end{equation}
where
\begin{eqnarray}
 & f_M(v) &= \frac{n_{e}}{\sqrt{2\pi T_e}}\, \exp{\left[ -\frac{(v-V_e)^2}{2T_e} \right]} \label{eq:fM} \\
 & f_b(v) &= \frac{n_b}{\sqrt{2\pi T_b}}\, \exp{\left[ -\frac{(v-V_b)^2}{2T_b} \right]} \label{eq:fB}
\end{eqnarray}

\begin{figure}[!htb]
\epsfxsize=8.5cm \centerline{\epsffile{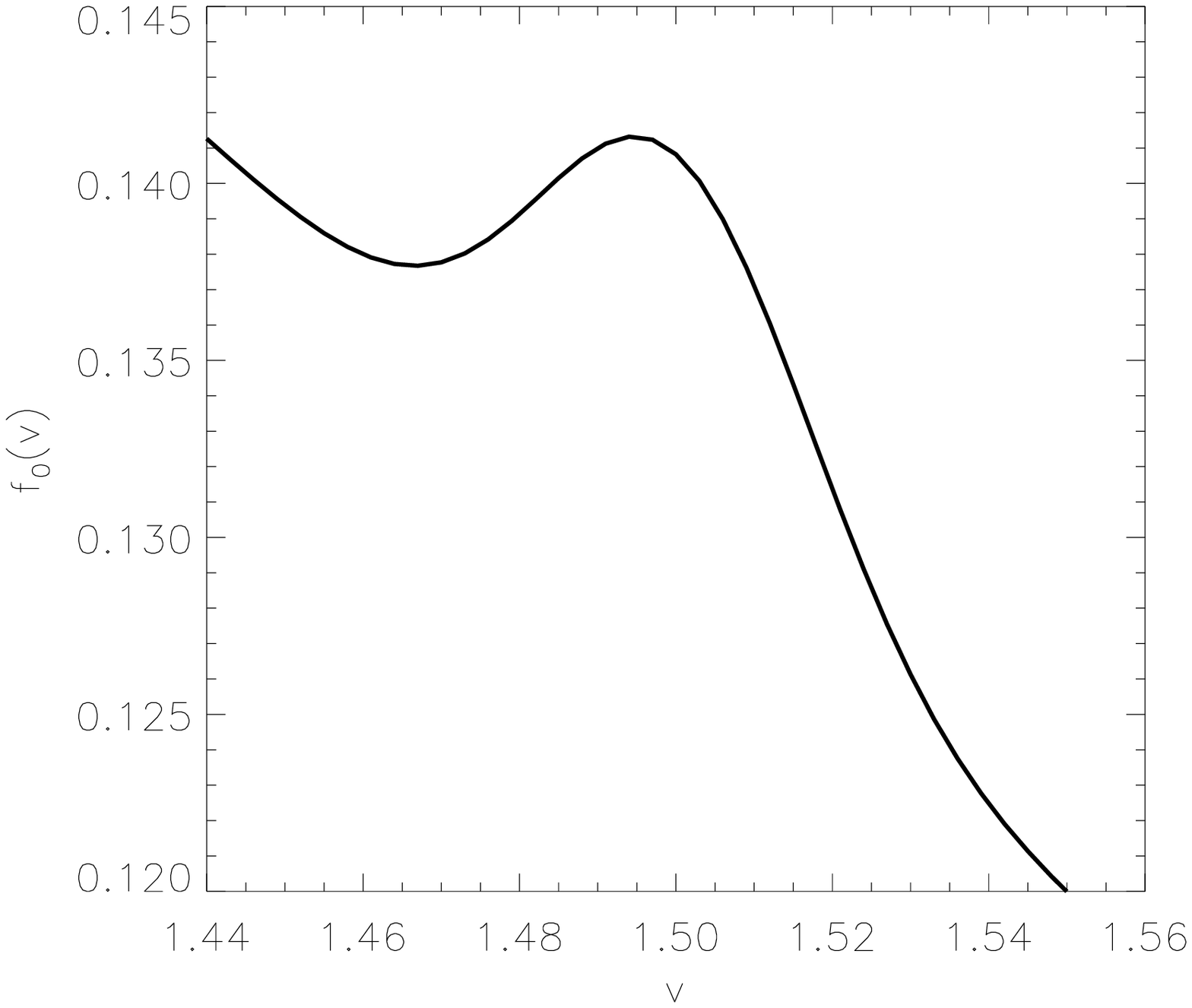}}   
\caption{Velocity profile of the equilibrium electron distribution function $f_0(v)$ around the beam speed $V_b$, for the case 
$V_b=1.5$. }
\label{fig1}
\end{figure}

Here, $n_{e(b)}$, $V_{e(b)}$ and $T_{e(b)}$ are  the density, bulk speed and temperature of the main (beam) electron population
and $n_e + n_b=n_0=1$. Moreover, in order to avoid a net electron current density which could, in principle, break the
electrostatic approximation, we imposed $n_e V_e+n_b V_b =0$. Typical parameters for the main and beam populations used for our
numerical study are: $n_b= 5\times 10^{-4}$, $T_e = 1$ and $T_b=3\times 10^{-4}$, while $V_b$ has been varied in the range
$0.7\le V_b \le 1.9$. The velocity profile of the equilibrium distribution function $f_0(v)$, zoomed in the velocity range around
the beam speed, is shown in Fig. \ref{fig1} for $V_b=1.5$; here, a small bump generated by the low density cold electron beam is
visible at $v=V_b$.

\begin{figure}[!htb]
\epsfxsize=8.5cm \centerline{\epsffile{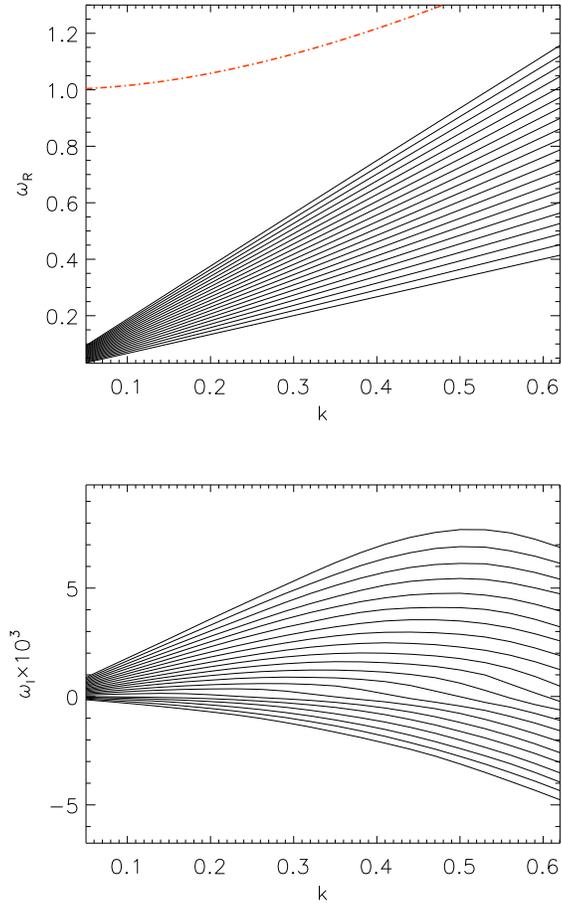}}   
\caption{Real part $\omega_R$ (top) and imaginary part $\omega_I$ (bottom) of the wave frequency as a function of the wave number
$k$, for different values of the beam speed $V_b$ ($V_b$ increasing from bottom to top curves). The red dot-dashed line in the top
plot indicates the curve for Langmuir waves.}
\label{fig2}
\end{figure}

In order to analyze the linear regime of wave propagation in presence of an electron beam, we employed a linear kinetic solver,
which computes numerically the Landau integral in the complex plane and solves for the roots of the dielectric function
\cite{oneil68}, when the equilibrium distribution function is chosen to be $f_0(v)$ in Eq. (\ref{eq:f0}). Figure \ref{fig2} shows
the dependence of the real part $\omega_R$ (top) and the imaginary part $\omega_I$ (bottom) of the wave frequency as a function of
the wavenumber $k$, for several values of $V_b$ ($V_b$ increases going from the bottom to the top curve). As a reference, the
dashed curve in Fig. \ref{fig2} (top) indicates the dispersion relation of Langmuir waves, which co-exist with beam modes for the
equilibrium distribution function $f_0(v)$.

In agreement with O'Neil and Malmberg \cite{oneil68}, the beam modes have an acoustic-like dispersion relation (see the top panel
of Fig. \ref{fig2}), with phase speed $v_\phi\lesssim V_b$. Moreover, by looking at the bottom panel of Figure \ref{fig2}, it is
evident that the beam modes are damped for small values of $V_b$ and can become unstable as $V_b$ increases.

\begin{figure}[!htb]
\epsfxsize=8.5cm \centerline{\epsffile{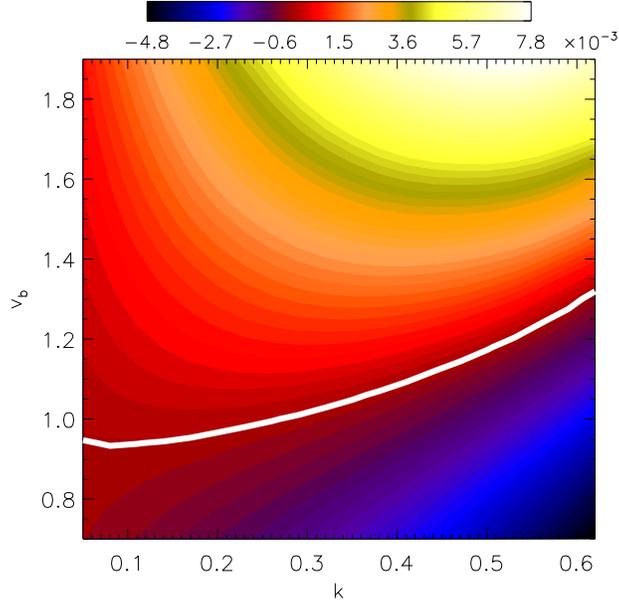}}   
\caption{Contour plot of $\omega_I$ in the $k-V_b$ plane. The white solid line corresponds to the curve $\omega_I(k,V_b)=0$.}
	\label{fig3}
\end{figure}

In Figure \ref{fig3}, we report the contour plot of $\omega_I$ in the $k-V_b$ plane. The white-solid line in Fig. \ref{fig3}
represents the curve $\omega_I\left(k,V_b\right)$=0, which marks the transition between stable and unstable regions. It is clear
from this plot that a large portion of the $k-V_b$ space is dominated by unstable solutions. 

\section{Numerical results}
\label{sec:numres}
For studying the nonlinear regime of the interaction between the plasma and the electron beam, we numerically integrated Eqs.
(\ref{eq:vlas}--\ref{eq:pois}) through an Eulerian code based on an explicit finite difference method for the evaluation of
spatial and velocity derivatives of the distribution function and on the well-known splitting scheme \cite{cheng76} for the
temporal evolution of $f$ (see Refs. \cite{pezzi13,pezzi14a} for details about the numerical algorithm). The numerical phase space
domain $\left[0, L\right] \times \left[-v_{max},v_{max} \right]$ is discretized with $N_x$ and $N_v$ grid points; here, $L$ is the
spatial length of the numerical box and $v_{max}=6v_{th,e}$. Periodic boundary conditions are imposed in physical
space, while, in velocity space, we set $f\left(x,|v|>v_{max},t \right)=0$. Since periodic boundary conditions are implemented in
physical space, Poisson equation is solved through a standard Fast Fourier Transform routine. Finally, the time step $\Delta t$
has been chosen in such a way to satisfy the Courant-Friedrichs-Levy \cite{peyret83} condition for the numerical stability of time
explicit finite difference schemes. In each numerical run described in the following, the initial equilibrium distribution
function is $f_0(v)$ in Eq. (\ref{eq:f0}) with $V_b=1.5$.

In this Section we discuss the results of two simulation runs: i) the propagation of a monochromatic beam wave in both linear
and nonlinear regimes and ii) the generation of isolated electrostatic structures produced by the nonlinear interaction of
beam-modes and associated with the coalescence of phase space vortices.

\subsection{Propagation of monochromatic beam modes in linear and nonlinear regime}
We perturb the initial equilibrium configuration described in the previous Section through a density perturbation of the form
$\delta n(x)=A_0 \sin(k_0 x)$, where $A_0=10^{-4}$, $k_0=2\pi/L=0.452$ and $L=13.89$. We follow the system evolution up to a
time $t_{max}=4000$. The number of grid points in physical and velocity domain is $N_x=256$ and $N_v=4001$, respectively. 

\begin{figure*}[!htb]
\epsfxsize=17cm \centerline{\epsffile{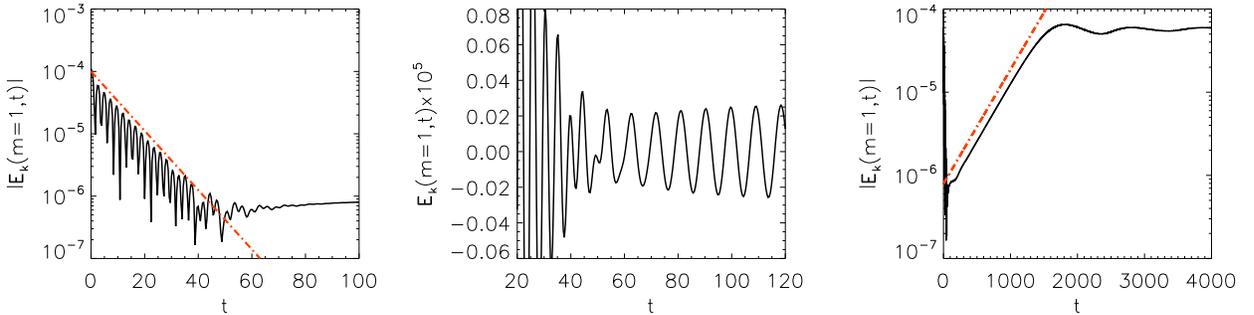}}  
\caption{On the left, semi-log plot of the time evolution of the electric Fourier component $m=1$ $|E_k(m=1,t)|$, in the early
times of the simulation. The red dot-dashed line represents the theoretical damping rate for Langmuir oscillations
$\omega^L_I=-0.101$. In the middle, time evolution of $E_k(m=1,t)$ in the time interval $20\leq t \leq 120$: a clear frequency
shift is observed here from the Langmuir frequency toward the frequency of the beam mode. On the right, time evolution of
$|E_k(m=1,t)|$ in semi-log plot for the total duration of the run. The red dot-dashed line represents the theoretical
prediction for the growth rate of the beam mode.} 
\label{fig4}
\end{figure*}

To compare with the linear expectations for the beam-modes described in Ref. \cite{oneil68} and revisited in Section
\ref{sec:lindisp}, we report, in Fig. \ref{fig4}, the time evolution of the fundamental electric field spectral component $E_k
(m=1)$ (corresponding to wavenumber $k_0$).

In the early stage of the simulation (left panel), one observes the exponential damping of the Langmuir oscillations initially 
excited by the  density perturbation. The red-dashed line in Fig. \ref{fig4} represents the theoretical prediction for the Landau
damping rate $\omega^L_I=-0.101$ for Langmuir waves, computed through the linear kinetic solver discussed previously. A very good
agreement can be seen here between the theoretical expectation and the observed damping. 

At $t\simeq50$, when the amplitude of the Langmuir oscillations has been decreased by about two orders of magnitude, the wave
damping effect saturates. In the middle panel of Fig. \ref{fig4}, reporting the electric signal evolution zoomed in the time
interval $t\in[20,120]$, a clear frequency shift can be noticed; from this time on, the electric oscillations occur with the
frequency of the beam mode, as predicted by the linear kinetic solver. 

When looking at the complete time evolution of $E_k(m=1,t)$, reported in the right panel of Fig. \ref{fig4}, it is evident that,
once the initial Langmuir oscillations have been completely dissipated, the amplitude of the remaining beam mode increases
exponentially. This is due to the fact that, with the parameters chosen for this simulation, the excited beam mode is unstable
(see Figure \ref{fig3} for $V_b=1.5$ and $k=0.452$). In fact, the red-dot-dashed line in this panel indicates the theoretical
prediction for the growth rate of the beam mode $\omega_I^b=0.0032$, as obtained through the linear kinetic solver; a very good
agreement between theoretical expectation and numerical results can be again appreciated here. Finally, in the late time regime,
the exponential growth of $E_k(m=1,t)$ saturates due to nonlinear effects and the wave amplitude oscillates around a nearly
constant value.

\begin{figure}[!htb]
\epsfxsize=8.5cm \centerline{\epsffile{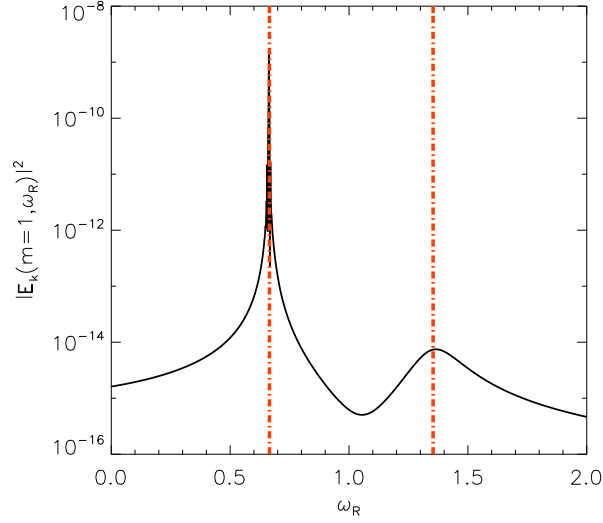}}   
\caption{Spectral electric energy $|E_k(m=1,\omega_R)|^2$ as a function of $\omega_R$. The red dot-dashed vertical lines 
indicate the frequency of the beam-mode $\omega_R^b=0.665$ and of the Langmuir wave $\omega_R^L=1.354$, respectively.} 
\label{fig5}
\end{figure}

In Figure \ref{fig5} we display the dependence of the spectral electric energy $|E_k(m=1,\omega_R)|^2$ on the real part of the
wave frequency $\omega_R$ (resonance curve). As the spectral energy has been computed by Fourier transforming the electric signal
over the entire spatio-temporal domain of the simulation, two peaks are visible in the resonance curve: the first dominant peak at
low frequency corresponds to the beam mode with $\omega_R^b\simeq 0.665$ ($v_\phi^b\simeq 1.47$), while the second one is
associated with the Langmuir oscillations, with $\omega_R^L=1.354$ ($v_\phi^L\simeq 2.99$), which are damped in the early stage of
the system evolution. The vertical red dot-dashed lines in this figure represent the frequencies of beam and Langmuir waves
respectively, computed through the linear kinetic solver.

\begin{figure*}[!htb]
\epsfxsize=17cm \centerline{\epsffile{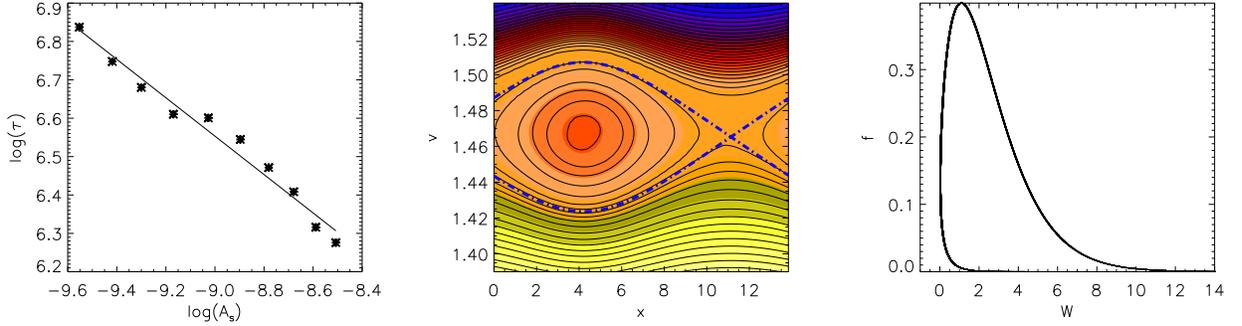}}  
\caption{On the left, logarithm of the characteristic oscillation period $\log{(\tau)}$ of the electric wave envelope in the
nonlinear phase (see right panel of Fig. \ref{fig4}) plotted versus the logarithm of the saturation amplitude $\log{(A_s)}$. The
dots are the numerical results, while the straight black line refers to the linear fit ($r=0.994$) given by Eq.
(\ref{eq:LinFitSat}). In the middle, phase space contour plot (and level lines) of the electron distribution function at
$t=t_{max}=4000$ and for $A_0=10^{-2}$. The blue dot-dashed curves are the separatrices  $v_\pm(x)=v_\phi \pm  
\sqrt{2(\phi_{max}+\phi(x, t_{max}))}$, where $\phi_{max}=\max_x\left(-\phi(x,t_{max})\right)$. On the right, scatter plot of the 
electron distribution function, at $t=t_{max}=4000$ and for $A_0=10^{-2}$, as a function of the energy in the wave frame.}
\label{fig6}
\end{figure*}

We move, now, to the analysis of the nonlinear evolution of the beam-mode instability observed in Fig. \ref{fig4}, by focusing in
particular on the nonlinear dynamics responsible for the saturation of the linear exponential growth. As demonstrated by O'Neil
in 1965 \cite{oneil65}, nonlinear effects come into play in the process of resonant wave-particle interaction at a time of
the order of the so-called trapping time $\tau_{trap}\approx (m_e/e kE)^{1/2}$, namely the  oscillation period of a charge in the
wave potential well. Following O'Neil, in nonlinear regime, due to the presence of a population of trapped particles, the
amplitude of an electrostatic wave oscillates in time over a period $\simeq\tau_{trap}$. Therefore, we performed several
simulations, by varying the amplitude of the initial density perturbation in the range $2.5 \times 10^{-3}\leq A_0 \leq 2.5 \times
10^{-2}$, and computed the characteristic oscillation time $\tau$ of the electric wave envelope (see right panel of Fig.
\ref{fig4} for $t>2000$). In Fig. \ref{fig6} (left) we report as dots the dependence of $\log{\tau}$ on $\log{A_s}$, $A_s$ being
the saturation amplitude of the electric oscillations in the nonlinear phase, evaluated as the average amplitude of the electric
envelope oscillations after the saturation of the linear instability. The black solid line in this panel represents the best fit
curve:
\begin{equation}
\label{eq:LinFitSat}
\log(\tau)=-(0.50\pm 0.03)\log(A_s)+(2.10\pm 0.26)
\end{equation}
As it is clear from this plot, the numerical dots show a clear linear trend, with a significantly high linear correlation
coefficient ($r=0.994$); the angular coefficient of the linear trend is $\alpha=-0.50\pm 0.03$, this meaning that $\tau$ is
inversely proportional to the square root of $A_s$, in good agreement with O'Neil's predictions.

The above considerations suggest that particle trapping is at play in the nonlinear phase of the system evolution. To further
confirm the existence of a population of trapped particles, in the middle panel of Fig. \ref{fig6} we show the phase space
contours of the electron distribution function at the end of the simulation and in the velocity range around $V_b$, for the
simulation with $A_0=10^{-2}$. The contour lines of the electron distribution function display a vortical structure, typical 
signature of particle trapping. Moreover, the blue-dashed lines in this panel represent the separatrix curves defined as 
$v_\pm(x)=v_\phi \pm  \sqrt{2(\phi_{max}+\phi(x, t_{max}))}$, where $\phi_{max}=\max_x\left(-\phi(x,t_{max})\right)$, which 
separate the phase space region with closed lines (trapped particles) from those with open lines (free particles). 

Finally, in the right panel of Fig. \ref{fig6} we plot the distribution function $f$ at the end of the run as a function of the
energy in the wave frame $W(x,v)=(v-v_\phi)^2/2-\phi(x, t_{max})$. That is, for each $(x,v)$ in the simulation domain, we plot
$f(x,v)$ vs $W(x,v)$, resulting in the single curve shown in Fig. \ref{fig6}. The fact that the points in this scatter plot fall
on a single curve shows that the electron distribution $f$ is a function of the energy $W$ alone, as expected for solutions of the
BGK type \cite{BGK}. From this evidence, we conclude that the final nonlinear state reached after the saturation of the linear
instability is, in fact, a BGK state.

\subsection{Nonlinear generation of isolated electrostatic pulses from the coalescence of phase space vortices}
In this Section we discuss the results of an additional simulation in which we impose a low amplitude random phase density noise
on the equilibrium distribution function in Eq. (\ref{eq:f0}), with $V_b=1.5$. The system evolution is followed up to a much
longer time with respect to the previous run ($t_{max} = 65000$). The parameters of this simulation are $n_e=0.995$, $T_e=1$,
$n_b=5\times10^{-3}$, $T_b=8\times10^{-4}$; in particular, we increased the beam density and decreased its temperature with
respect to the previous run. Moreover, $L=500$ ($k_0=2\pi/L=0.013$), $N_x=2048$ and $N_v=4001$.

The random phase density disturbance is composed of $128$ Fourier modes and has the form $\delta n (x) = A_0 \,g(x)/\max_x[g(x)]$,
where $A_0=10^{-6}$ and 
\begin{equation}
g(x) =  \sum_{j=1}^{128} sin(jk_0x+\varphi_j)  \, ,
\label{eq:gx}
\end{equation}
being $\{\varphi_j\}$ random phases in $[0,2\pi]$.

\begin{figure}[t]
\epsfxsize=8.5cm \centerline{\epsffile{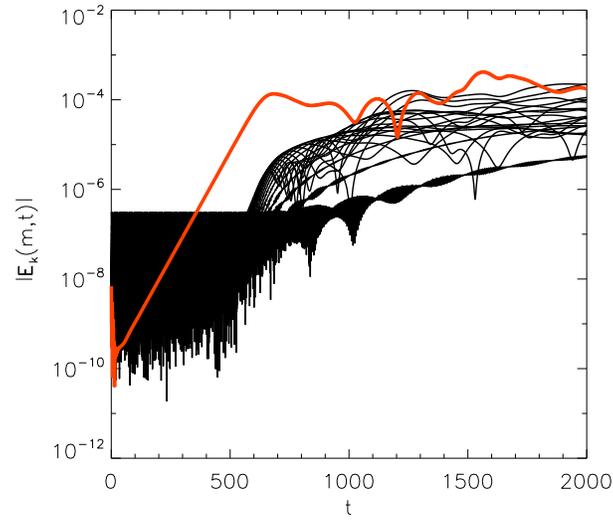}}   
\caption{Time evolution of the electric field spectral components $1\leq m \leq 20, m=45$, in semi-log plot. The red solid curve
represents the most unstable mode, $m=45$.}
\label{fig7}
\end{figure}

Fig. \ref{fig7} reports the time evolution of $E_k(m,t)$, in the time interval $t\leq 2000$, for several spectral components
($1\leq m \leq 20$ and $m=45$). The most unstable Fourier component, $m=45$, is plotted in red. The time evolution of this
component shows a preliminary damping of the initial Langmuir wave, as discussed in previous Section, and then a linear
instability phase, followed by a saturation in nonlinear regime. It is worth noting from this plot that the other Fourier
components start growing in amplitude once the most unstable mode has reached its saturation level, this suggesting that a
nonlinear secondary instability took place to excite the whole Fourier spectrum.

\begin{figure}[t]
\centering
\epsfxsize=8.5cm \centerline{\epsffile{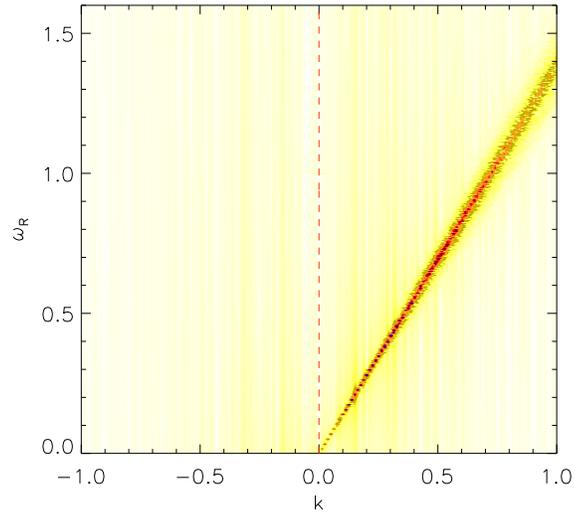}}  
\caption{$\omega_R-k$ Fourier spectrum of the electric signal, taken over the whole spatial domain and in the time interval
$0\leq t\leq 5000$.}
\label{fig8}
\end{figure}

To analyze the characteristic of the wave signals shown above, in Fig. \ref{fig8} we report the $\omega_R-k$ Fourier spectrum of
the electric field. As shown in this plot, an acoustic branch has been excited, whose phase speed is $v_\phi=1.40$, in good
agreement with the prediction of the linear kinetic solver for the beam mode branch with the parameters used in this simulation. 
It is interesting to note in this plot that: (i) since the Fourier spectrum has been taken over a long time interval ($t<5000$),
no signature of Langmuir activity is recovered, as Langmuir fluctuations have been rapidly Landau damped after few wave cycles in
the early stage of the system evolution; (ii) the beam fluctuations propagate only along the positive $x$ direction (positive
values of the wavenumber), accordingly to the direction of propagation of the electron beam in the equilibrium distribution
function [see Eq. (\ref{eq:f0})].

\begin{figure*}[t]
\epsfxsize=14.5cm \centerline{\epsffile{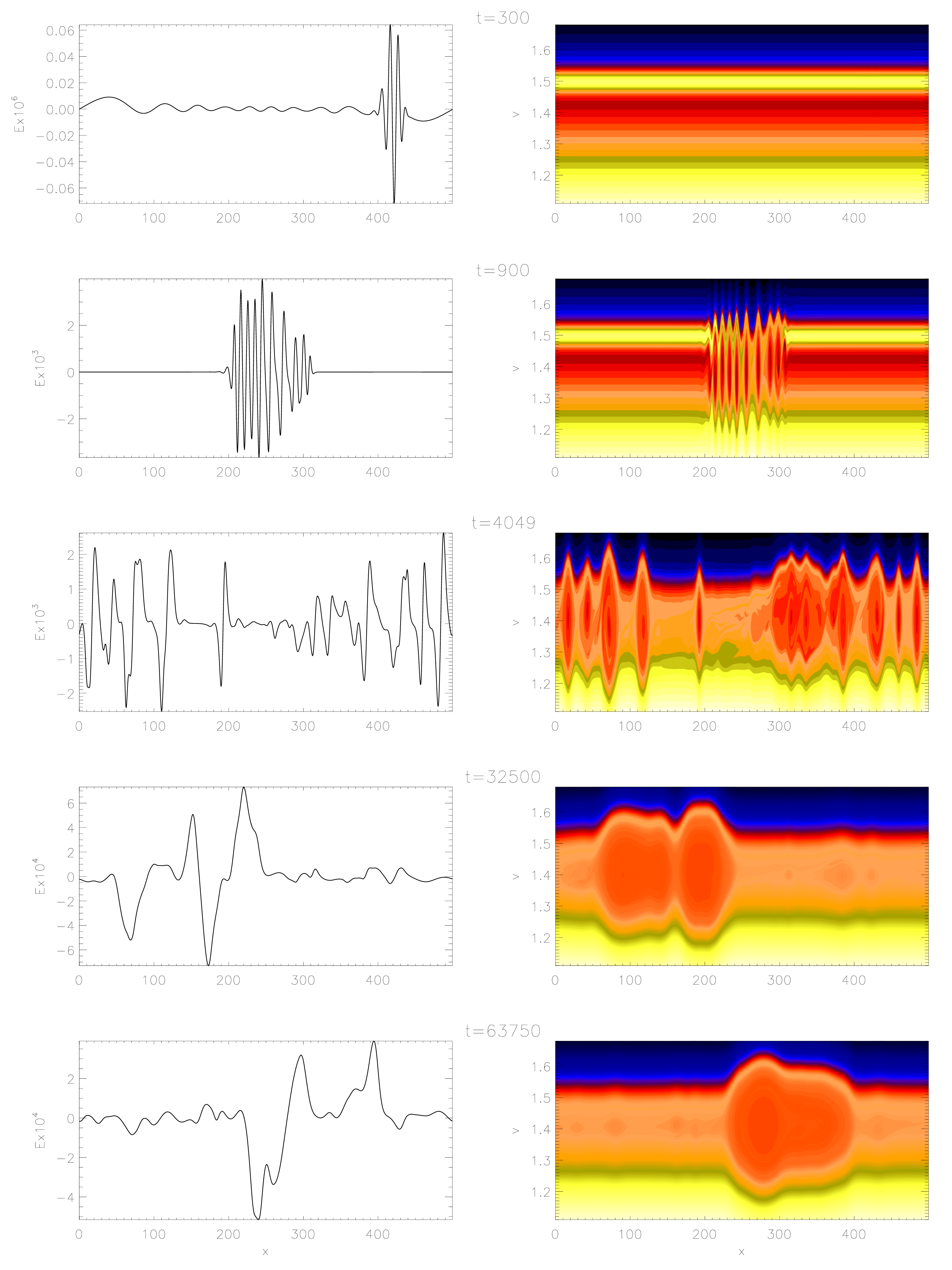}}   
\caption{Spatial profile of the electric field at different time instants in the simulation (left column) and the corresponding
phase space contour plot of the electron distribution function (right column), in a velocity range around the drift velocity of
the beam $V_b$.}
\label{fig9}
\end{figure*}

At this point, it is interesting to look at the evolution of the phase space distribution function, as the linear instability
develops, with the most unstable mode growing, and later on once the nonlinear regime has been reached and a significant energy
has been transferred to high wavenumber spectral components. In Fig. \ref{fig9} we report the electric field signals (left column)
at different times in the simulation and the corresponding phase space portraits of the electron distribution function (right
column). At $t=300$ (top panels), in the middle of the linear instability phase, the electric field has the form of a small
amplitude localized wave packet, corresponding to the most unstable wavenumber ($m=45$); here, the amplitude of the electric field
is too small to produce any significant deformation of the distribution function, whose level lines appear indeed unperturbed. 

At $t=900$, after the nonlinear saturation of the instability, many spectral components gained significant energy and the electric
field amplitude has rapidly increased of many orders of magnitude; now, particles are trapped in the wave potential and vortical
structures appear in phase space. At later times ($t=4049$), the whole Fourier spectrum has been excited, high wavenumber
components have energy comparable to the linearly most unstable mode and the electric field signal appears as a train of solitary
structures, propagating along the positive $x$ direction; to each electric field pulse corresponds a phase space vortex in the
electron distribution function. These phase space vortices tend to merge and coalesce and, as a consequence, the number of
electric pulses decrease in time ($t=32500$). Finally, in the large-time regime ($t=63750$), the last couple of phase space
structures are coalescing, showing a tendency towards the formation of a single phase space vortex with a corresponding
soliton-like electric field signal. We point out that the process of vortex merging occur on very slow time scales, whose final
time asymptotic configuration is presumably reached in times beyond the maximum time of our simulation.

\section{Summary and conclusions}
\label{sec:concl}
In this paper we analyzed through a numerical approach the interaction of a beam of electrons with a background globally
neutral plasma, both in linear and nonlinear regime. This problem has been already analytically approached by O'Neil and Malmberg
in 1968 \cite{oneil68} in linear regime. These authors pointed out that the presence of the electron beam generates an additional
wave branch of the electrostatic dielectric function, with acoustic dispersion relation, that they named beam modes.

In the first part of the present work, we revisited the results in Ref. \cite{oneil68} by means of a linear kinetic solver, which
computes numerically the roots of the plasma dielectric function. The results of this preliminary linear analysis are in good
agreement with Ref. \cite{oneil68}, confirming that the so-called beam modes have linear dependence of their frequency
on the wavenumber and phase speed close to the beam speed. We also pointed out that, depending on the characteristics of the
electron beam (density, temperature and speed), the beam modes can become unstable. As summarized in Fig. \ref{fig3}, it is
possible to excite beam modes in a large portion of the $k-V_b$ space and most of the time the beam mode oscillations are
unstable. We decided, therefore, to study in detail the excitation process of these beam modes, as well as the linear and
nonlinear phase of their instability, since the interaction of a plasma with a beam is a relevant physical situation, frequently
encountered both in space and laboratory plasma systems.

To do so, we employed a numerical Vlasov-Poisson solver, which allows to follow the evolution of the electron distribution
function in 1D-1V phase space. We performed two different numerical experiments, in which the equilibrium configuration,
consisting in a background of Maxwellian electrons and the electron beam (plus motionless neutralizing protons) is perturbed by
small amplitude density disturbances; in both simulations the typical parameters of the electron beam and of the initial
perturbation are chosen in such a way to trigger the beam mode instability.

In the first simulation, we perturbed the equilibrium configuration with a monochromatic density perturbation. This simple
situation allowed us to analyze in detail first of all the linear regime of the system evolution, making contact with the
predictions of the linear kinetic solver: very good agreement has been found between numerical results and predictions of the
linear solver concerning both real and imaginary part of the wave frequency. Furthermore, we focused on the nonlinear saturation
of the beam-mode instability, showing that the nonlinear regime of wave propagation is associated with the generation of a trapped
particle population and, in general, to a phase space distribution function of the BGK type. 

For the second simulation, we launched the electron beam in an initially Maxwellian plasma with random density noise. By looking
at the time evolution of several electric field Fourier components, we observed the exponential growth of the amplitude of the
most unstable one and then its nonlinear saturation, as in the previous run. However, once this mode has reached its nonlinear
saturation amplitude, other Fourier components start gaining energy, until the whole Fourier spectrum is excited, following
the acoustic-like dispersion relation of the beam modes. In the late time regime, the resulting spatial profile of the electric
field appears as a train of isolated pulses; a phase space vortex in the electron distribution function is associated with each of
these pulses. These vortices in phase space tend to merge and the electric pulses to collapse correspondingly: this dynamical
evolution is expected to result in a single long-lived phase space structure, which propagates in time keeping unchanged its speed
and shape (the corresponding electric signal displaying a soliton-like profile 
\cite{mangeney99,malaspina13,pezzi14b}). 

We emphasize that the generation of these electrostatic pulses is a pure nonlinear kinetic effect, as it is related to the kinetic
processes of phase space vortex generation and merging.

\section*{Acknowledgements}
This work has been supported by the Agenzia Spaziale Italiana under the Contract No. ASI-INAF 2015-039-R.O ``Missione M4
di ESA: Partecipazione Italiana alla fase di assessment della missione THOR.''

\clearpage

\begin{thebibliography}{99}
\bibitem{oneil68}T.M. O'Neil and J.H. Malmberg, Phys. Fluids  {\bf 11}, 1754--1760, (1968).
\bibitem{landau46} L.D. Landau, Zh. Eksp. Teor. Fiz. {\bf 10}, 25 (1946).
\bibitem{oneil65} T.M. O'Neil, Phys. Fluids {\bf 8}, 2255--2262 (1965).
\bibitem{manfredi97} G. Manfredi, Phys. Rev. Lett. {\bf 79}, 2815 (1997).
\bibitem{brunetti00} M. Brunetti, F. Califano and F. Pegoraro, Phys. Rev. E {\bf 62}, 4109 (2000).
\bibitem{galeotti05} L. Galeotti and F. Califano, Phys. Rev. Lett. {\bf 95}, 015002 (2005).
\bibitem{valentini05} F. Valentini, V. Carbone, P. Veltri and A. Mangeney, Phys. Rev. E {\bf 71}, 017402 (2005).
\bibitem{califano06} F. Califano, L. Galeotti and A. Mangeney, Phys. Plasmas {\bf 13}, 082102 (2006).
\bibitem{galeotti06} L. Galeotti, F. Califano and F. Pegoraro, Phys. Lett. A {\bf 355}, 381 (2006).
\bibitem{danielson04} J. R. Danielson, F. Anderegg and C.F. Driscoll, Phys. Rev. Lett. {\bf 92}, 245003 (2004) 
\bibitem{holloway91} J.P. Holloway and J.J. Dorning, Phys. Rev. A {\bf 44},3856 (1991).
\bibitem{valentini06} F. Valentini, T.M. O’Neil, and D.H.E. Dubin, {\it Phys. Plasmas} {\bf 13}, 052303 (2006).
\bibitem{BGK} I.B. Bernstein, J.M. Greene, and M.D. Kruskal, {\it Phys. Rev.} {\bf 108}, 546 (1957).
\bibitem{kabantsev06} A.A. Kabantsev, F. Valentini and C.F. Driscoll, Non-Neutral Plasma Physics VI {\bf 862}, 13--18 (2006).
\bibitem{anderegg09a} F. Anderegg, C.F. Driscoll, D.H.E. Dubin and T.M. O'Neil, Phys. Rev. Lett. {\bf 102}, 095001 (2009).
\bibitem{anderegg09b} F. Anderegg, C.F. Driscoll, D.H.E. Dubin, T.M. O'Neil and F. Valentini, Phys. Plasmas {\bf 16}, 055705 
(2009).
\bibitem{afeyan04} B. Afeyan, K. Won, V. Savchenko, T.W. Johnston, A. Ghizzo, and P. Bertrand, in {\it Proceedings of the 3rd 
International Conference on Inertial Fusion Sciences and Applications M034, Monterey, CA, 2003}, edited by B. Hammel et al. 
(American Nuclear Society, LaGrange Park, IL, 2004), p. 213.
\bibitem{johnston09} T. W. Johnston, Y. Tyshetskiy, A. Ghizzo, and P. Bertrand, Phys. Plasmas {\bf 16}, 042105 (2009).
\bibitem{valentini12} F. Valentini, D. Perrone, F. Califano, F. Pegoraro, P. Veltri, P.J. Morrison and T. M. O’Neil, Phys. 
Plasmas {\bf 19}, 092103 (2012).
\bibitem{marsch06} E. Marsch, Living Rev. Sol. Phys. {\bf 3(1)}, 1--100 (2006).
\bibitem{valentini11a} F. Valentini, F. Califano, D. Perrone, F. Pegoraro and P. Veltri, Phys. Rev. Lett. {\bf 106}, 165002 
(2011).
\bibitem{valentini11b} F. Valentini, F. Califano, D. Perrone, F. Pegoraro and P. Veltri, Plasma Phys. Control. Fusion {\bf 53}, 
105017 (2011).
\bibitem{vecchio14} A. Vecchio, F. Valentini, S. Donato, V. Carbone, C. Briand, J. Bougeret and P. Veltri, J. Geop. Res. {\bf 
119(9)}, 7012--7024 (2014).
\bibitem{valentini14b} F. Valentini, A. Vecchio, S. Donato, V. Carbone, C. Briand, J. Bougeret and P. Veltri, The Astrophys. 
Journal Letter {\bf 788}, L16 (2014).
\bibitem{cheng76} C. Cheng and G. Knorr, J. Comput. Phys. {\bf 22.3}, 330--351 (1976).
\bibitem{pezzi13} O. Pezzi, F. Valentini, D. Perrone \& P. Veltri, Phys. Plasmas {\bf 20}, 092111 (2013).
\bibitem{pezzi14a}O. Pezzi, F. Valentini, D. Perrone \& P. Veltri, Phys. Plasmas {\bf 21}, 019901 (2014). 
\bibitem{peyret83} R. Peyret and T. D. Taylor, {\it Computational Methods for Fluid Flow} (Springer, New York, 1983).
\bibitem{mangeney99} A. Mangeney, C. Salem, C. Lacombe, J.L. Bougeret, C. Perche, R. Manning, P.J. Kellogg, K. Goetz, S.J. Monson,
J.M. Bosqued, Ann. Geophys. {\bf 17}, 307 (1999)
\bibitem{malaspina13} D.M. Malaspina, D.L. Newman, L.B. Willson, K. Goetz, P.J. Kellogg, K. Kerstin, J. Geophys. Res. Space Phys.
{\bf 118}, 591 (2013).
\bibitem{pezzi14b} O. Pezzi, F. Valentini and P. Veltri, Eur. Phys. J. D {\bf 68}, 128 (2014).
\end{thebibliography}
\end{document}